\documentclass[sigconf]{acmart}

\usepackage{soul}
\usepackage[inline]{enumitem}
\usepackage{url}
\usepackage{multirow}
\usepackage{makecell}
\usepackage{pifont}
\usepackage{subcaption}
\usepackage{cleveref}

\newcommand{\floor}[1]{\left\lfloor #1 \right\rfloor}

\newcommand{\xmark}{\ding{55}}%

\AtBeginDocument{%
  \providecommand\BibTeX{{%
    \normalfont B\kern-0.5em{\scshape i\kern-0.25em b}\kern-0.8em\TeX}}}

\copyrightyear{2021}
\acmYear{2021}
\setcopyright{acmcopyright}
\acmConference[MM '21]{Proceedings of the 29th ACM International Conference on Multimedia}{October 20--24, 2021}{Virtual Event, China}
\acmBooktitle{Proceedings of the 29th ACM International Conference on Multimedia (MM '21), October 20--24, 2021, Virtual Event, China}
\acmPrice{15.00}
\acmDOI{10.1145/3474085.3479236}
\acmISBN{978-1-4503-8651-7/21/10}

\begin{document}
\fancyhead{}

%% Use this when submitting an article to a sponsored event. You'll
%% receive a unique submission ID from the organizers
%% of the event, and this ID should be used as the parameter to this command.
%%\acmSubmissionID{123-A56-BU3}

%%
%% The majority of ACM publications use numbered citations and
%% references.  The command \citestyle{authoryear} switches to the
%% If you are preparing content for an event
%% sponsored by ACM SIGGRAPH, you must use the "author year" style of
%% the next command will enable that style.

%%
%% end of the preamble, start of the body of the document source.

% \title{DEPA: SELF-SUPERVISED AUDIO EMBEDDING FOR DEPRESSION DETECTION}
\title{DEPA: Self-Supervised Audio Embedding for Depression Detection}


\author{Pingyue Zhang}
\affiliation{%
  \institution{MoE Key Lab of Artificial Intelligence\\ 
  X-LANCE Lab\\
  Department of Computer Science and Engineering\\
  Shanghai Jiao Tong University}
  \city{Shanghai}
  \country{China}
}
\email{williamzhangsjtu@sjtu.edu.cn}

\author{Mengyue Wu$\dag$}
\thanks{$\dag$Mengyue Wu and Kai Yu are the corresponding authors.}
\affiliation{%
  \institution{MoE Key Lab of Artificial Intelligence\\ 
  X-LANCE Lab\\
  Department of Computer Science and Engineering\\
  Shanghai Jiao Tong University}
  \city{Shanghai}
  \country{China}
}
\email{mengyuewu@sjtu.edu.cn}

\author{Heinrich Dinkel}
\authornote{This author is currently affiliated with Xiaomi Tech. Ltd., Beijing}
\affiliation{%
  \institution{MoE Key Lab of Artificial Intelligence\\ 
  X-LANCE Lab\\
  Department of Computer Science and Engineering\\
  Shanghai Jiao Tong University}
  \city{Shanghai}
  \country{China}
}
\email{heinrich.dinkel@gmail.com}

\author{Kai Yu$\dag$}
\affiliation{%
  \institution{MoE Key Lab of Artificial Intelligence\\ 
  X-LANCE Lab\\
  Department of Computer Science and Engineering\\
  Shanghai Jiao Tong University}
  \city{Shanghai}
  \country{China}
}
\email{kai.yu@sjtu.edu.cn}

\renewcommand{\shortauthors}{Zhang and Wu, et al.}

\begin{abstract}
Depression detection research has increased over the last few decades, one major bottleneck of which is the limited data availability and representation learning.
Recently, self-supervised learning has seen success in pretraining text embeddings and has been applied broadly on related tasks with sparse data, while pretrained audio embeddings based on self-supervised learning are rarely investigated.
This paper proposes \textit{DEPA}, a self-supervised, pretrained \textit{dep}ression \textit{a}udio embedding method for depression detection.
An encoder-decoder network is used to extract DEPA on in-domain depressed datasets (DAIC and MDD) and out-domain (Switchboard, Alzheimer's) datasets.
With DEPA as the audio embedding extracted at response-level, a significant performance gain is achieved on downstream tasks, evaluated on both sparse datasets like DAIC and large major depression disorder dataset (MDD).
This paper not only exhibits itself as a novel embedding extracting method capturing response-level representation for depression detection but more significantly, is an exploration of self-supervised learning in a specific task within audio processing.
\end{abstract}


% \ccsdesc[500]{Computing methodologies~Neural networks}
% \ccsdesc[500]{Computing methodologies~Supervised learning by classification}
% \ccsdesc[500]{Computing methodologies~Supervised learning by regression}





\maketitle

\section{Introduction}
Depression, a disease of considerable attention, has been affecting more than 300 million people worldwide. 
An increasing amount of research has been conducted on automatic depression detection and severity prediction, in particular, from conversational speech, which has embedded crucial information about one's mental state. 
Despite recent advances in deep learning, automatic depression detection from speech remains a challenging task. 

Since depression is a complicated mental disorder consisting of various symptoms, utilizing traditional feature extraction methods for emotion recognition might lack precision in asserting each individual's mental state. 
Previous exploration has covered commonly-used emotion-related features such as COVAREP~\cite{degottex_covarep_2014}, general-purpose audio features including log-Mel spectrogram (LMS), and combination of Short-Time Fourier Transform (STFT) and Mel-Frequency Cepstral Coefficients (MFCC)~\cite{rejaibi2019mfcc}, and speaker-related audio embedding like i-vector~\cite{Cummins2014}. 
However, as these features are not tailored for application in assessing mental disorders, thus could be less efficient towards such a task with high specificity.  
%Previous speech-based detection work has experimented on various acoustic features, including prosodic features (e.g., pitch, jitter, loudness, speaking rate, energy, pause time, intensity), spectral features (e.g., formants, energy spectrum density, spectral energy distribution, vocal tract spectrum, spectral noise) and cepstral features (e.g., Mel-Frequency Cepstral Coefficients (MFCC)~\cite{rejaibi2019mfcc,Salekin2018}), and more recently, feature combinations like COVAREP (CVP), which consists of a high-dimensional feature vector covering common features such as fundamental frequency and peak slope.

Another characteristic of depression is that usually only one single label (diagnosis results) is provided for a multi-turn interview. 
Specifically, during a session with a doctor, it would be impossible to give a specific label $y_t \in \{0,1\}$, representing the mental state (depressed or healthy) for each time step $t$.
Here $t$ can be chosen on any arbitrary level, such as phone-, word- or sentence-level.
Those long sequences subsequently influence depression detection performance.
Previous work has hinted that extracting embeddings on segment-level (e.g., sentence, response) might benefit performance~\cite{rejaibi2019mfcc}, while modeling depression via a stationary, time-step independent representation is likely to fail~\cite{AlHanai2018}.
Hence, a successful audio-embedding for depression detection needs to be extracted on sequence-level (e.g., spoken sentence/utterance), to capture rich, long-term spoken context as well as emotional development within an interview.

%TODO Previous representation learning has covered 
%Many diseases, mental or physical, usually take time to develop, meaning that pin-pointing the exact time at which a person has contracted a specific disease is nearly impossible.
%Depression detection is an example of such a disease, which at its core is a weakly supervised problem. 
%Only a single label (severity) is provided as a result of a multi-turn interview.
The last important problem is that models so far are heavily restricted by the limited amount of depression data. 
Hence even with the recent advances of deep learning, this data sparsity has caused difficulty in model performance enhancement and reproduction.
One potential solution to the aforementioned data sparsity problem is to pretrain a model on large data and then leverage the model's knowledge to a downstream task.
However, pretraining on supervised tasks (e.g., speech recognition) is time-intensive and costly with manual labeling.
Self-supervised training, which utilizes the original property of the data, can potentially remove the dependency on manual labels, thus being able to easily scale with data.

\begin{figure}[htbp]
    \centering
    \includegraphics[width=\linewidth,scale=0.25]{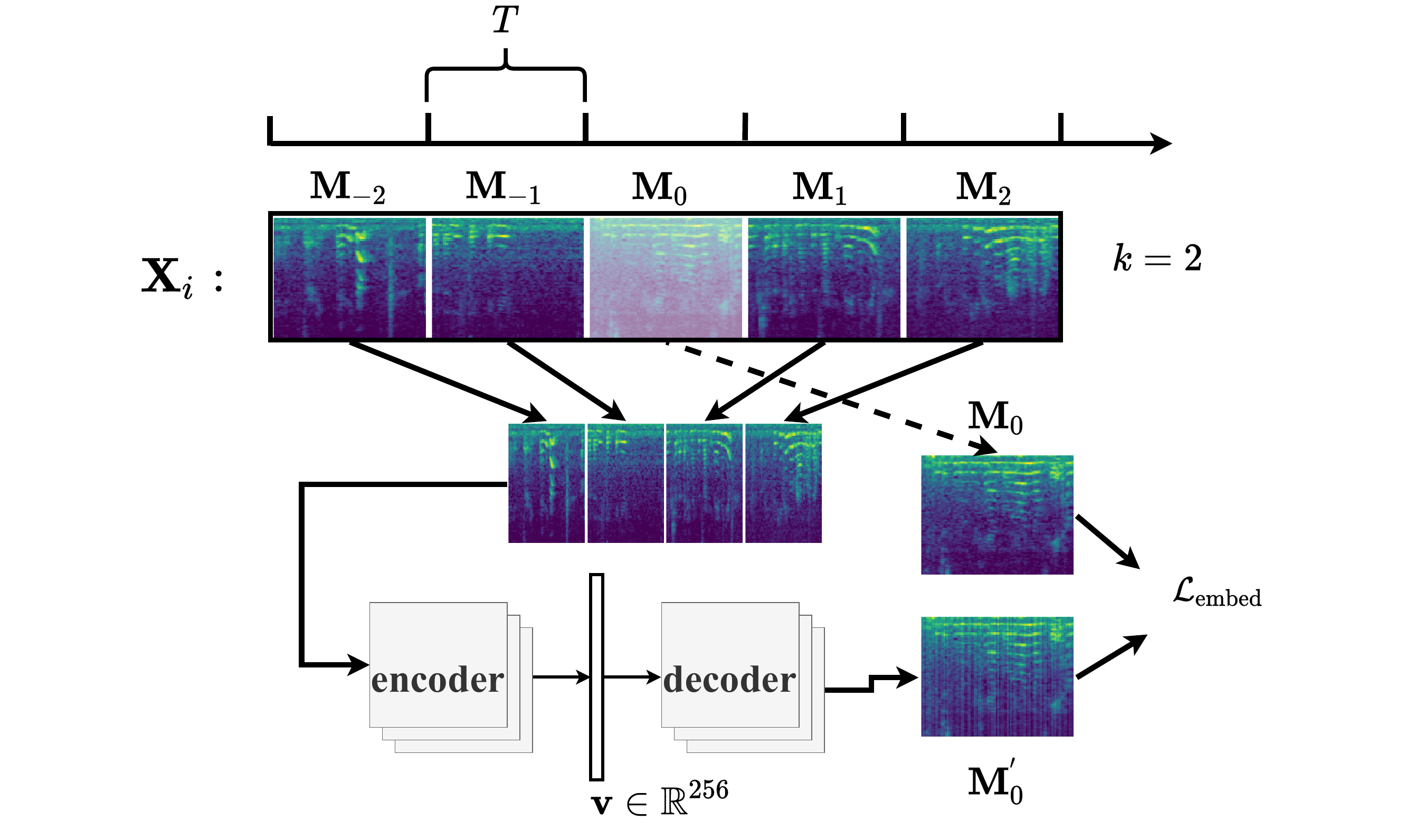}
    \caption{The \textit{DEPA} pretraining framework. The training objective follows the estimation of a middle spectrogram within a sequence of $2k+1$ spectrograms.}
    \label{fig:pretrain}
\end{figure}

\paragraph*{Contribution} This paper proposes \textit{DEPA}, a self-supervised, pretrained \textit{dep}ression \textit{a}udio embedding method for automatic depression detection (see \Cref{fig:pretrain}).
%and ameliorate the previously mentioned data sparsity problem. 
To our knowledge, this is the first time a self-supervised neural network pretraining is performed on a depression detection task. 
\begin{itemize}
    \item We achieved the highest classification performance and lowest regression errors on the benchmark depression detection dataset by modeling each patient's response via the sequence of his/her uttered speech and realize the extraction of response-level representation with DEPA. 
    \item To highlight the necessity of using sentence-level representations for tasks like depression detection, we compared with previously-used audio features, including general-purpose features, emotion-related representations, and x-vector speaker embeddings. Results suggest a significant performance gain with the use of DEPA and the efficiency of sequence-level representations. We also design several experiments to further illustrate performance enhancement by using sentence-level representations.
    \item DEPA pretrained on depression data (in-domain) and out-domain datasets are compared, including other mental disorders interviewing conversation datasets and general-purpose speech datasets. Results indicate that self-supervised pretraining on large datasets, especially on those share similar mental disorders with depression, is beneficial to the current data sparsity scenarios and largely outperforms raw features without pretraining.
    \item We conduct a series of ablation studies to analyze possible factors which may influence the pretraining process, including the configuration of feature extraction, the hyperparameters in pretraining process, and the pretraining strategy. 
\end{itemize}

\section{Related Work}
\label{sec:related_work}

In this section, related work on depression detection and self-supervised learning will be discussed. 
\subsection{Depression detection}
Various methods have been proposed for automatic depression detection. 
Representation learning and classifier selection are the two major research areas within depression detection.
Deep learning methods have been employed to extract high-level feature representations~\cite{AlHanai2018}.
In particular~\cite{haque2018measuring} utilized causal convolutional neural networks (C-CNN) to enable sequence-level feature extraction and achieved a high performance by combining visual, audio, and textual modalities. 
Results indicate that sequence-level representation outperforms frame-level ones with respect to depression detection. 
%It has been observed that low-level audio features are in terms of performance inferior to high-level text-based features.
Notable work on pretraining audio features for depression detection includes~\cite{Salekin2018}, which trained an audio word-book in unsupervised fashion using Gaussian mixture models to extract segment-level features and uses a BLSTM model with max-temporal pooling as a depression classifier.
\cite{rejaibi2019mfcc} investigated a knowledge transfer from emotion recognition to depression detection by firstly pretraining a recurrent neural network on a fully-labeled emotion recognition dataset.
Their results suggest that emotion is a possible marker for automatic depression detection and that transfer learning enhances performance.

%Despite the tryout on different features~\cite{avec2016_valstar,Salekin2018} and models, the predicted severity scores from speech-based depression detection is average.
%Work in~\cite{dinkel2019text} indicated that large data, task-independent, text-embedding pretraining, can significantly enhance depression detection performance. 

\subsection{Self-supervised learning}

Self-supervised learning is a technique where training data is autonomously labeled, yet the training procedure is supervised.
A classic example of self-supervised learning is auto-encoders~\cite{10.1007/978-3-030-26061-3_30}, aiming to reconstruct a given input from a hidden representation.
Learning representations with self-supervised training has lead to remarkable improvement in several fields, including textual, visual, audio, and multimodal processing. 
In natural language processing (NLP), self-supervised text embedding pretraining can be seen as a major breakthrough, with methods such as GloVe~\cite{pennington2014glove}, BERT~\cite{devlin2018bert}, and ELMo~\cite{Peters:2018}.
Self-supervised pretraining from audio-visual signals, such as SoundNet~\cite{aytar2016soundnet}, have been found to outperform traditional spectrogram-based features in acoustic environment classification.
In fact, much research has focused on self-supervised audio-visual segmentation and feature extraction~\cite{Rouditchenko2019,Zhao2018,owens2018audio}.
Recently, many researches in Computer Vision field use contrastive learning as their self-supervised learning method. SimCLR~\cite{chen2020simple}, MoCo~\cite{he2020momentum}, and CoCLR~\cite{han2020self} all use contrastive learning with some variations to retain visual representations.

In particular, pretrained approaches such as EmoAudioNet~\cite{Rejaibi2019} have been applied in depression detection, however, its pretraining process requires large 1000h (Librispeech), gender-labeled training data to be successful, which requires extensive manual labor. 
%Under this circumstance, labeling those quantities of data is costly.
Our main inspiration for this work stems from Audio2Vec~\cite{tagliasacchi2019self}, where a self-supervised approach was proposed, the objective of which is to extract general-purpose audio representations for mobile devices.

Relatively, little research has been conducted on self-supervised audio representation learning in depression detection, or such medical applications. 
The reasons could include:
\begin{enumerate*}[label=\itshape\arabic*\upshape)]
    \item Content-rich audio contains undesirable information, such as environmental sounds, interfering speech, and noise.
    \item Features are typically low-level and extracted within a short time-scale (e.g., 40 ms), each containing little information about high-level concepts (e.g., a single phoneme contains little information about a sentence).
    \item Due to the nature of depression detection, interviews are comparatively long (many minutes), which, combined with fine-scale features, means that a classifier needs to remember very long sequences while filtering out unimportant information.
\end{enumerate*}

\section{DEPA: Self-Supervised Audio Embedding}
\label{sec:self_supervised_pretraining}
This paper proposes DEPA, an auditory feature extracted via a neural network to summarize spoken language.
Our proposed method consists of a self-supervised convolutional encoder-decoder network, where the encoder is later used as DEPA embedding extractor from spectrograms.
Given a spectrogram of a specific audio clip (e.g., a spoken sentence) $\mathbf{X} \in \mathbb{R}^{S\times D}$, where $S$ is the number of frames and $D$ the data dimension (e.g., frequency bins).
We proceed to slice $\mathbf{X}$ into $\floor{\frac{S}{(1+\alpha)(2k+1)\cdot  T}}$ non-overlapping samples $\mathbf{X}_i \in \mathbb{R}^{((2k+1) \cdot T) \times D}$, where $T$ is the number of frames in one sub-spectrogram which will be explained below, $k$ is the hyperparameter which controls number of such sub-spectrograms in one sample, and $\alpha \geq 0$ is the gap parameter such that each sample being $gap \in \mathbb{R}^{(\alpha \cdot (2k+1) \cdot T)\times D}$ apart.
The gap between two segments avoids that the self-supervised model exploits spectral leakage to shortcut and easily solve the task. 

$$
\mathbf{X} = \left[\mathbf{X}_0, gap, \mathbf{X}_1, gap, \cdots, \mathbf{X}_i, \cdots, \right]
$$

Each sample $\mathbf{X}_i$ is sliced into $2k + 1$ sub-spectrograms $\mathbf{M}_j \in \mathbb{R}^{T \times D}$. Each $\mathbf{X}_i$ is therefore the concatenation of a center $\mathbf{M}_0$ sub-spectrogram and its $k$ adjacent left-right context:
% Each $\mathbf{M}_i$ contains $k$ of them before and after a center one $\mathbf{M}_0$: 
$$
\mathbf{X}_i = \left[\mathbf{M}_{-k}, \cdots, \mathbf{M}_{-1}, \mathbf{M}_{0}, \mathbf{M}_{1}, \cdots,  \mathbf{M}_{k}\right],
$$

If $\mathbf{X}$ is shorter than $(2k+1)T$ frames, we pad zeros to fill $(2k+1)T$ frames.

\begin{figure}[htbp]
\centering
\includegraphics[width=0.8\linewidth,scale=0.25]{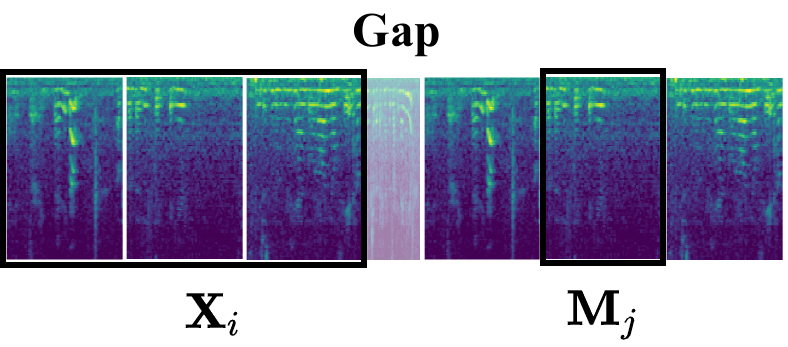}
\caption{Slicing an audio clip spectrogram ($\mathbf{X}$) into samples $\mathbf{X}_i$ and sub-spectrograms $\mathbf{M}_j$ for DEPA training. The gap avoids spectral leakage of a sub-spectrogram to its neighbors.}
\label{fig:slice}
\end{figure}

Our self-supervised learning uses a generative strategy. The training process treats the center spectrogram $\mathbf{M}_0$ as the target, taking its surrounding spectrograms $\mathbf{M}_j, (j\neq 0)$ to re-generate the center spectrogram and computes the embedding loss (\Cref{eq:selfloss}). \Cref{fig:slice} shows the slicing process, and the detailed pretraining process is depicted in \Cref{fig:pretrain}.

% The self-supervised training process treats the center spectrogram $\mathbf{M}_0$ as the target label, given its surrounding spectrograms $\mathbf{M}_j, (j\neq 0)$ and computes the embedding loss (\Cref{eq:selfloss}). \Cref{fig:slice} shows the slicing process, and the detailed pretraining process is depicted in \Cref{fig:pretrain}.

% Loss of this encoder-decoder network mean squre error (MSE): 
\begin{equation}
\label{eq:selfloss}
% \mathcal{L} =\frac{1}{n} \sum_{i=1}^n 
\mathcal{L}_{embed} = \frac{1}{TF} \sum_{t=1}^{T}\sum_{d=1}^{F} (\mathbf{M}_{0_{t,d}} - \mathbf{M}_{0_{t,d}}^{'})^2.
\end{equation}\\

\paragraph*{Encoder architecture} The encoder architecture contains three downsampling blocks, followed by an extra convolution layer as well as an adaptive pooling layer.
Each block consists of a convolution, average pooling, batch-normalization, and rectified linear unit (ReLU) activation layer. 
The time-axis $2kT$ is subsampled to $\frac{2kT}{64}$ before being average pooled in time and frequency dimension.

\begin{figure}[htbp]
\centering
\includegraphics[width=\linewidth]{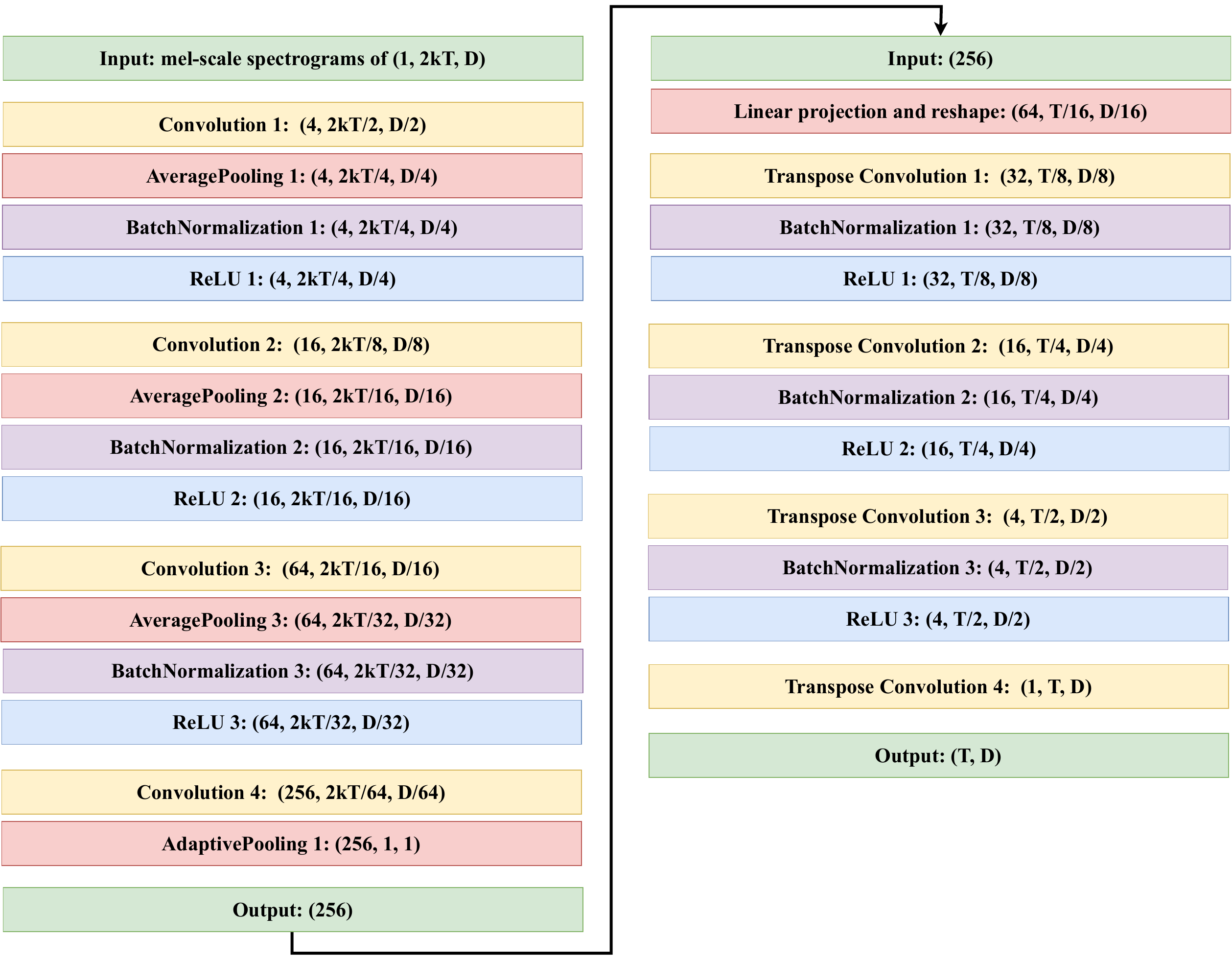}
\caption{DEPA pretraining encoder-decoder architecture.}
\label{fig:encoder-decoder}
\end{figure}

\paragraph*{Decoder architecture}
The decoder upsamples the encoder output $\mathbf{v}$ via four transposed convolutional upsampling blocks and predicts the center spectrogram $\mathbf{M}_0^{'} \in \mathbb{R}^{T \times D}$.
% The model is then updated via the embedding loss in \Cref{eq:selfloss}.
The encoder-decoder architecture is shown in \Cref{fig:encoder-decoder}.

After pretraining the encoder-decoder network, DEPA is extracted via feeding a variable-length audio segment $\mathbf{R}$ (here on patient response-level) into the encoder model and obtaining a single $|\mathbf{v}|=256$-dimensional embedding per segment.
The sequence of DEPA embeddings is then further fed into a depression detection network, which can be seen in \Cref{fig:detection_model}.

\section{Downstream Task: Depression Detection}
\label{secc:detection_model}
In this section, we detail our approach to the downstream task of depression detection on two datasets: DAIC, a small dataset used as depression detection benchmark; MDD, a large dataset focused specifically on female patients with major depression detection (see \cref{subsec:dataset} for a detailed introduction). 
%In this section, we detail our approach to two different downstream tasks (DAIC, MDD) of depression detection after DEPA extraction.
% \paragraph*{BLSTM Model with Multi-Task Learning}
% The final decision about the depression state and severity is carried out by a multi-task training scheme, combining depression state classification and depression score prediction.
\paragraph*{Small Benchmark Data with Two Label Sets}
Depression state and severity score is provided in the DAIC dataset, hence, we propose a multi-task scheme, combining depression state classification and depression score prediction.
This approach models a patients' depression sequentially, meaning that only the patients' responses are utilized.
Due to the recent success of LSTM networks in this field~\cite{AlHanai2018}, our depression prediction structure follows a bidirectional LSTM (BLSTM) approach with four layers of size $128$.

\begin{figure}[htbp]
\centering
\includegraphics[width=0.89\linewidth]{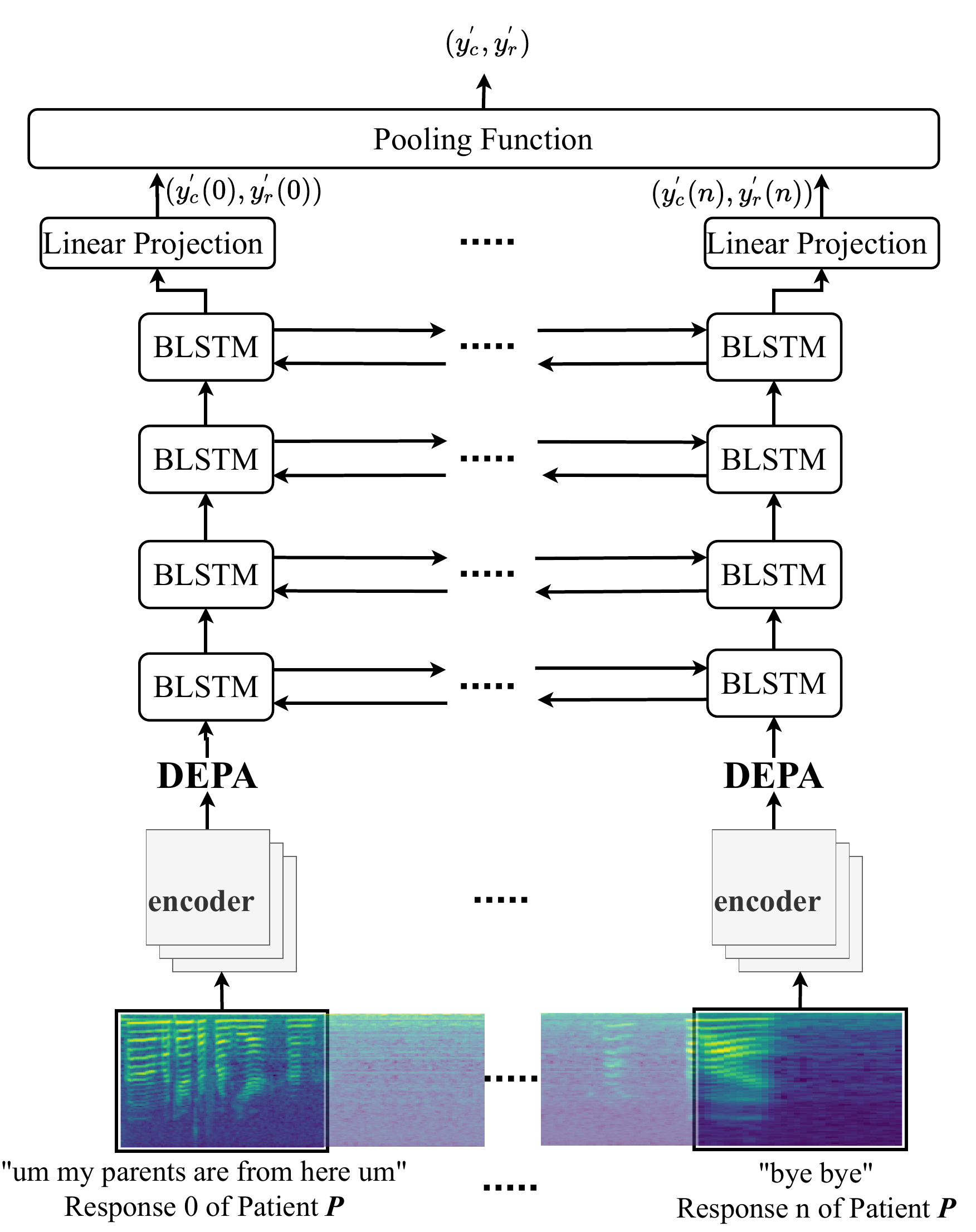}
\caption{Depression detection with DEPA on DAIC with multi-task training scheme. The encoder from the proposed encoder-decoder model provides the BLSTM network with high-level auditory features. In this figure, DEPA is extracted on response-level.}
%Weighted average pooling via attention is utilized.}
\label{fig:detection_model}
\end{figure}

The model outputs for response $r$ a two dimensional vector $({y_{c}^{'}}(r), {y_{r}^{'}}(r))$, representing the estimated binary patient state ($y_{c}^{'}(r)$) as well as the PHQ-8 score ($y_{r}^{'}(r)$, a numerical metric to evaluate depression extent).
Finally, first timestep pooling is applied to compile all responses of a patient to a single vector $({y_{c}^{'}}(0), {y_{r}^{'}}(0))$. 
The architecture is shown in \Cref{fig:detection_model}.

\begin{align}
    \ell_{bce}(y_c^{'}, y_c) &= \text{-}[y_c \cdot \log y_c^{'}+(1 - y_c) \log (1- y_c^{'})] \label{eq:bce} \\
    \ell_{hub}(y_r^{'}, y_r) &=
        \begin{cases}
        0.5 (y_r- y_r^{'})^2, & \text{if } |y_r- y_r^{'}| < 1 \\
        |y_r- y_r^{'}| - 0.5, & \text{otherwise }
        \end{cases}\label{eq:huber}\\
    \ell(y_c^{'}, y_c ,y_r^{'}, y_r ) &= \ell_{bce}(\sigma(y_c^{'}), y_c) + \ell_{hub}(y_r^{'}, y_r)
\end{align}

Two outputs are constructed, one directly predicts the binary outcome of a participant being depressed, the other outputs the estimated PHQ-8 score. 
We opt to use a combination of binary cross entropy (BCE, for binary classification, \Cref{eq:bce}) and Huber loss (for regression, \Cref{eq:huber}). 
$y_c, y_r$ are the ground truth binary and PHQ-8 score, respectively, while $\sigma$ is the sigmoid function.
In this way, our model considers the internal relationship between binary classification and PHQ-8 score regression, where a higher PHQ-8 score commonly indicates a probability of being classified as depressed.

\paragraph*{Large Data with One Classification Label}
MDD, a privately collected large depression dataset, is also applied in our downstream detection task. For this dataset, we merely predict the depression state, which is the only label provided. Therefore, the utilized method is similar to the one above with minor changes: the BLSTM only output one scaler: $y_c^{'}$, and only a binary cross-entropy loss $\ell_{bce}$ is used. 
Similarly, we model the patient's response in a sequential manner.

\section{Experimental Setup}
\label{sec:experiments}
\paragraph*{Depression Data}
\label{subsec:dataset}
A commonly used dataset within depression detection is the Distress Analysis Interview Corpus -- Wizard of Oz (DAIC)~\cite{DeVault:2014:SKV:2615731.2617415} dataset, which encompasses 50 hours of data collected from a total of 142 patients.
%The DAIC database was previously used for the Audio/Visual Emotion Challenges.
Two labels are provided for each participant: a binary diagnosis of depressed/healthy and the patient's eight-item \textbf{P}atient \textbf{H}ealth \textbf{Q}uestionnaire score (PHQ-8) metric. 
Thirty speakers within the training (28 \%) and 12 within the development (34 \%) set are classified to have depression. 
The DAIC dataset is fully transcribed, including corresponding on- and offsets within the audio.
While this dataset contains training, development, and test subsets, our evaluation protocol is reported on the \textit{development} subset, since test subset labels are only available to participants of the 2017 Audio/Visual Emotion Challenge (AVEC). 

\begin{table}[htbp]
\centering
\begin{tabular}{l|r|r|r|r}
Dataset &      &   Train & Dev & Test \\
\hline\hline
\multirow{3}{*}{DAIC} &  \#(D/H) &  30/77 &  12/23 &  - \\
& \# responses &  158 &  190 &  - \\
& $\varnothing$ response (s) &  2.74 &  2.63 &  - \\
\hline
\multirow{3}{*}{MDD} &  \#(D/H) &  516/357 &  101/87 &  105/83 \\
& \# responses &  318 &  320 &  373 \\
& $\varnothing$ response (s) &  0.77 &  0.769 &  0.78 \\
\end{tabular}
\caption{Statistics regarding the number of responses for each subset. D/H represents depressed and healthy patients respectively.}
\label{tab:statistics_responses}
\end{table}

\begin{table}[htbp]
    \centering
    \begin{tabular}{l|r|r|r}
        Domain & Dataset & Duration (h) & Language   \\
        \hline\hline
        \multirow{2}{*}{In} & DAIC & 13 & English \\
        & MDD & 411 & Mandarin\\
        \hline
        \multirow{2}{*}{Out} & SWB & 300 & English \\
        & AD & 400 & Mandarin\\
    \end{tabular}
    \caption{In- and out-domain datasets used for DEPA pretraining.}
    \label{tab:dataset}
\end{table}

In addition, a large conversational dataset (MDD) for major depression disorder detection under collection has now consisted of 1000 hours of speech conversation between interviewers and subjects, with a balanced proportion of healthy and depressed participants (722 depressed and 527 healthy). We split the dataset into a training set ($70\%$), a development set ($15\%$), and a test set ($15\%$). Unlike the fully-transcribed DAIC dataset, no annotation is provided in MDD. 
We hence applied the X-vector-based speaker diarization tool provided by the Kaldi Toolkit~\cite{povey2011kaldi} to extract all patient's speaking segments from the audio.
MDD is incorporated to highlight the benefit of summarizing long sequences using DEPA. 
Detailed statistics regarding the proportion of depressed/healthy subjects, the number of patient responses, and their average duration is displayed in \Cref{tab:statistics_responses}.

%Note that most of the experiments and ablation study are conducted on DAIC dataset because it's more popular. 

\begin{table*}[htbp]
    \centering
    \begin{tabular}{llll|rrr|rr}
    \multicolumn{4}{c|}{} & \multicolumn{3}{c|}{Classification} & \multicolumn{2}{c}{Regression}\\
    \hline
    Method &  \makecell[c]{Feature} & \makecell[c]{Feature\\level} & \makecell[c]{Pretrain \\ Database}& Pre & Rec & F1 & MAE & RMSE \\
     \hline\hline
    %\xmark & \xmark & MSP &0.71 & 0.53 & 0.62 & 6.07 & 6.94\\
    \cite{williamson2016detecting} & dMFCC-VT & Frame & \xmark & - & - & 0.57 & 5.32 & 6.38 \\ % 1Loudness 2s / 1s | dMFCC segments with 1s -> 1ss
    \cite{ma2016depaudionet} & LMS & Frame & \xmark & 0.68 & 0.77 & 0.72 & - & - \\ % zhenyi 32ms , wind_size  64ms
%     \xmark &  & Alhanai et al.\cite{al2018detecting} &0.71 & 0.83 & 0.77 & 5.10 & 6.37\\
    \cite{AlHanai2018} & HCVP & Response & \xmark  & 0.71 & 0.56 & 0.63 & 5.13 & 6.50\\ % Response -level
    \cite{Stepanov2018} & LLD & Response & \xmark  & - & - & - & 4.96 & 6.32 \\ % % 20ms 10ms, Each turn , 24 functionsals for each turn 
    \cite{avec2016_valstar} & CVP & Frame & \xmark & 0.63 & 0.69 & 0.66 & 5.36  & 6.74 \\ % % 30 ms / 10 ms
    \cite{Rejaibi2019} & STFT+MFCC & Frame & LibriSpeech & - & - & 0.66 & - & - \\ % IS actually 12102102 500ms zhenyi 2.5s chuangkou
    \hline
    \multirow{4}{*}{BLSTM (Ours)} &  HCVP & Response & \xmark  & 0.73 & 0.66 & 0.69 & 4.95 & 6.45\\
     &  LMS & Frame & \xmark  & 0.61 & 0.61 & 0.61 & 5.68 & 6.51  \\ %43 Hz
     &  STFT & Frame & \xmark  & 0.64 & 0.64 &0.64 & 6.83 & 9.27  \\ %43 Hz
     &  x-vector & Response & Voxceleb  & 0.59 & 0.59 & 0.59 & 6.23 & 7.10  \\ %43 Hz
    %BLSTM &  STFT
     \hline
    \multirow{8}{*}{BLSTM (Ours) + DEPA} & LMS & \multirow{2}{*}{Response} & \multirow{2}{*}{DAIC}  & 0.71 & 0.65 & 0.68 & 5.47 & 6.33\\
    % \multirow{6}{*}{BLSTM (Ours) + DEPA} & LMS & Response & \multirow{2}{*}{DAIC}  & 0.64 & 0.66 & 0.63 & 5.60 & 6.47\\
     & STFT &  &  & 0.91 & 0.89 & 0.90 & 5.48 & 6.31\\
     %& STFT & Response &  & 0.746 & 0.746 & 0.746 & 5.01 & 6.07\\
     \cline{2-9}
     & LMS & \multirow{2}{*}{Response} & \multirow{2}{*}{MDD} & 0.75 & 0.74 & 0.75 & 5.10 & 6.05 \\
     &  STFT &  &  & \textbf{0.94} &  0.94 & \textbf{0.94} & 5.59 & 6.46\\
     \cline{2-9}
     & LMS & \multirow{2}{*}{Response} & \multirow{2}{*}{SWB} & 0.84 & 0.87 & 0.86 & 5.43 & 6.41\\
     &  STFT &  &  & 0.91 &  0.90 & 0.91 & 5.15 & 6.02\\
     \cline{2-9}
     & LMS & \multirow{2}{*}{Response} & \multirow{2}{*}{AD}  & 0.67 & 0.67 & 0.67 & 5.37 & 6.50\\
    %  & LMS & Response & \multirow{2}{*}{AD}  & 0.69 & 0.57 & 0.40 & 5.42 & 6.77\\
     &  STFT &  &   & 0.93 & \textbf{0.96} & \textbf{0.94} & \textbf{4.75} & \textbf{5.73}\\
     \hline
%     DAIC & mel DEPA & 0.72 & 0.72 & 0.72 & 4.72 & 6.10\\
%     % DAIC & STFT &0.76 & 0.78 & \bf{0.77} & 5.25 & 6.25\\
%     SWB & DEPA &0.80 & 0.69 & \bf{0.74} & 5.27 & 6.53\\
%     AD & DEPA &0.75 & 0.73 & \bf{0.74} & \bf{4.65} & \bf{5.99}\\
%      \hline
%     $\Sigma$ & DEPA &0.68 & 0.56 & 0.61 & 5.46 & 6.85 \\
    \end{tabular}
    \caption{Comparison between DEPA and other audio-based depression detection methods on the DAIC development set.}
%     \caption{Comparison between detection with and without DEPA pretraining, regarding three utlilzed datasets. $\Sigma$ represents the use of all three datasets for DEPA extraction.}
    \label{tab:result}
\end{table*}

\paragraph*{Pretraining Data}
We aim to compare DEPA in regards to pretraining on related, e.g., in-domain (depression detection) and out-domain (e.g., speech recognition) datasets.

Regarding in-domain data, we utilized the aforementioned DAIC and MDD datasets (we take a subset of 411 hours) for in-domain pretraining in order to compare DEPA to traditional audio feature approaches. 
In order to ascertain DEPAs' usability, we further used the mature Switchboard (SWB)~\cite{godfrey1992switchboard} dataset, containing 300 hours of English telephone speech.
% A large conversational dataset (MDD) for major depression disorder detection under collection has now consisted of 1000 hours speech conversation between interviewers and subjects, with balanced proportion of healthy and depressed participants (708 depressed and 512 healthy). 
Lastly, we utilized the Alzheimer's disease (AD) dataset, collected by Shanghai Mental Clinic Center~\cite{xiali2019}, containing about 400 hours (questions and answers) of Mandarin interview recordings from elderly participants. 
The four datasets are described in \Cref{tab:dataset}.

\paragraph*{Feature Selection}

Regarding front-end features, our work investigates common LMS and log-power STFT features.
Due to different sample rates across the datasets, we resample each dataset's audio to 22050 Hz.
All following features are extracted as default with a hop length ($\omega_{hop}$) of $5 ms$ and a Hann window length ($\omega_{win}$) of four times $\omega_{hop}$ (e.g., $20 ms$).
128 dimensional LMS and 512-dimensional STFT features were chosen as the default signal-processing front-end.
In order to compare DEPA against non-self-supervised approaches, $553$-dimensional higher-order (mean, median, variance, min, max, skewness, kurtosis) COVAREP~\cite{degottex_covarep_2014} (HCVP) features were extracted on response-level.
HCVP can be seen as a traditional high-level representation, which is an ensemble of lower-level descriptors, such as MFCC, pitch, glottal flow, and other features.

Lastly, we also extracted 256-dimensional x-vectors using a Resnet34 structure~\cite{snyder2018x}, for comparison purposes.
X-vectors, which are a state-of-the-art method within speaker recognition, have been seen to outperform traditional i-vectors, reported as markers for depression and some other mental diseases~\cite{Cummins2014}.
%Our utilized x-vector system using a Resnet34 structure has been trained on the VoxCeleb dataset~\cite{Nagraniy2017}.

\paragraph*{DEPA Pretraining Process}

Our encoder-decoder training utilizes LMS and STFT front-end features, with hyper-parameters $k=5, T=96, \alpha=0.1$, extracting a $|\mathbf{v}|=256$ dimensional DEPA embedding.
Moreover, the model is trained for $25$ epochs using Adam optimization with a starting learning rate of $0.004$, and a batch size of 512.
% As a preprocessing step, the AD dataset has been cut into segments ($\mathbf{X}$) via energy-based voice activity detection (VAD), since it contained very long utterances of \textgreater 30 min.
% Note that VAD is not applied for DAIC and SWB data since their audio segments are fully labeled or short in nature.
% The pretraining process differs for in-domain and out-domain datasets.
% For in-domain data, all responses of a patient are concatenated, meaning that silence or speech of the interviewer is neglected.
% For out-domain data, no preprocessing is done, meaning that the entire dataset is utilized.

%\subsection{Depression Detection}
%  ADD LEVEL ?

\paragraph*{Depression Detection Training Process}
As mentioned, for the DAIC dataset, we used a multi-task learning strategy to output both binary classification and PHQ-8 scores with a BLSTM network structure.
Regarding the MDD dataset, the BLSTM only output classification prediction. 
Data standardization was applied by calculating a global mean and variance on the training set and using those on the development set.
A dropout of $0.1$ was applied after each BLSTM layer to prevent overfitting.
Adam optimization with a starting learning rate of $4\mathrm{e}^{-5}$ and a batch size of 1 was used.

\paragraph*{Metrics}
Following previous work~\cite{AlHanai2018}, results are reported in terms of mean average error (MAE) and root mean square deviation (RMSE) for regression and macro-averaged (class-wise) precision, recall, and their harmonic mean (F1) score for classification.

\section{Results}
\label{sec:results}

%Baseline和实验结果还是放同一张表格，方便比较

% \begin{table}[h]
%     \centering
%     \begin{tabular}{rr|lll|ll}
%     \multicolumn{2}{c}{} & \multicolumn{3}{|c|}{Classification} & \multicolumn{2}{|c}{Regression}\\
%     \hline
%      Pretrain & Feature & Pre & Rec & F1 & MAE & RMSE \\
%      \hline\hline
%     \xmark & MSP &0.71 & 0.53 & 0.61 & 6.07 & 6.94\\
%     \xmark & HCVP &0.73 & 0.66 & 0.69 & 4.95 & 6.45\\
%      \hline
%     DAIC & DEPA & 0.72 & 0.72 & 0.72 & 4.72 & 6.10\\
%     % DAIC & STFT &0.76 & 0.78 & \bf{0.77} & 5.25 & 6.25\\
%     SWB & DEPA &0.80 & 0.69 & \bf{0.74} & 5.27 & 6.53\\
%     AD & DEPA &0.75 & 0.73 & \bf{0.74} & \bf{4.65} & \bf{5.99}\\
%      \hline
%     $\Sigma$ & DEPA &0.68 & 0.56 & 0.61 & 5.46 & 6.85 \\
%     \end{tabular}
%     \caption{Comparison between detection with and without DEPA pretraining, regarding three utlilzed datasets. $\Sigma$ represents the use of all three datasets for DEPA extraction.}
%     \label{tab:result}
% \end{table}
Results on the two different datasets are provided respectively: DAIC, a benchmark dataset for depression detection, is used to compare against previously methods and demonstrate how DEPA can help boost performance on sparse data scenarios; MDD, by contrast, provides insight on how DEPA compares with raw features, along with a different number of input responses. 

\subsection{DAIC Results}
Our results using the proposed BLSTM approach with and without DEPA pretraining are compared to previous attempts in \Cref{tab:result}. 
The results are analyzed on multiple levels.
\paragraph*{Feature Level Comparison}
The results in \Cref{tab:result} are in line with our initial assumption, that frame-level audio-features are indeed underperforming compared to response-level ones, especially for the BLSTM model.
This is likely due to the models' inherent incapability to remember very long sequences (\textgreater 10000 frames) for an abstract task such as depression detection, which is also commonly seen within other audio processing tasks where long sequences are harder to predict.
%This behavior is commonly seen within other tasks such as automatic speech recognition (ASR), where long utterances are generally harder to predict.
Regarding the classification results, it can be seen that traditional HCVP features outperform LMS, STFT, and x-vector approaches.
Specifically, with respect to regression, HCVP achieves a score of 4.95, much lower than any other frame-level feature approach (LMS, STFT, dMFCC-VT, CVP).
The sub-optimal performance of the x-vector system is likely due to the short response durations in this dataset, being on average $\approx$ 2 seconds long.
%(see \Cref{fig:train_dev_duration_dist}).
The performance of x-vector systems generally decreases for utterances shorter than 3 seconds~\cite{snyder2018x}.
Furthermore, response-level features are likely to contain more context-related information, while frame-level features tend to isolate information between frames.

\paragraph*{Feature Comparison} Even though a multitude of features are compared (MFCC, LMS, LLD, CVP, STFT), no clear trend can be established between feature and final performance.
Regarding our BLSTM approach, STFT features consistently underperform against LMS and HCVP features in terms of MAE.
This is likely due to the increased amount of parameters needed to be estimated by the BLSTM model (input layer increases from 128 to 512) in conjunction with the limited available training data.
This can partially be improved by either reducing the feature size (e.g., utilize LMS, MFCC features) or the number of samples per speaker (e.g., use the response averaged HCVP features). 
By contrast, when experimented with DEPA features, extracting DEPA from STFT features constantly outperforms DEPA from LMS features. %can be seen for all experiments.
%This is likely due to the embedding loss $\mathcal{L}_{embed}$ being more difficult for large dimensional features, thus aiding the model at learning.

\paragraph*{Pretraining Datasets Comparison} 
DEPA pretraining on the same DAIC dataset can be seen to enhance performance for LMS (F1 0.61 $\rightarrow$ 0.68) and especially STFT features (F1 0.64 $\rightarrow$ 0.90).
This, in turn, reinforces our initial assumption that response-level features are much more useful for depression detection.
Pretraining on large datasets (MDD, SWB, and AD) outperformed DAIC in terms of binary classification as well as regression.
Further, pretraining on AD resulted in the best performance in terms of all metrics.
Larger datasets (DAIC \textless SWB \textless MDD = AD) for DEPA pretraining generally improve performance for STFT features.

\subsection{MDD Results}
\begin{figure}[htbp]
\centering
\includegraphics[width=0.89\linewidth]{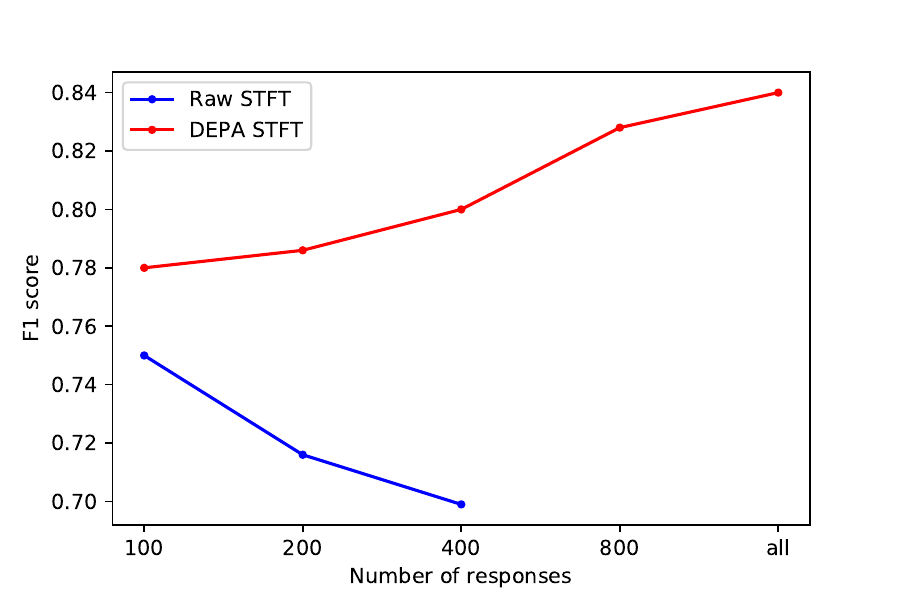}
\caption{Depression detection w and w/o DEPA pretraining on MDD, experimented with different number of patient's responses incorporated.}
%Weighted average pooling via attention is utilized.}
\label{fig:results_MDD}
\end{figure}

We further evaluated DEPA performance on large depression dataset MDD and reported the result on the test set, with the best setting observed from the DAIC experiments. 
We compared the difference of using raw STFT features and DEPA STFT pretrained on MDD, along with a different number of patient's speech responses as the model input. 
As seen in \Cref{fig:results_MDD}, DEPA largely outperforms raw features without DEPA pretraining, regardless of the input response numbers. We have not conducted experiments of raw STFT using more than 400 responses because the sequences are too long.
% Larger amounts of frames increase the performance, whereas the performance peaks when DEPA is extracted at the response level (the default).
% Further, DEPA exhibits its superior capability of summarizing long sequences, as the $F1$ score soars to 0.84 when we input all segments.
% By contrast, regarding raw features without DEPA, the performance is negatively correlated with the input segments. 
More input responses increase the performance and the $F1$ score soars to 0.84 when we input all responses with DEPA. 
On the contrary, when using raw features the performance is negatively correlated with the input responses.
The average duration of one response is around $0.7s$ (\Cref{tab:statistics_responses}) and raw STFT is extracted with a hop length $\omega_{hop}=5ms$. 
Hence one response approximately corresponds to a $140 \times 512$ raw STFT feature, which is 140 times longer than DEPA. 
This suggests DEPA's superior capability of summarizing sequences, leading to a performance enhancement with increasing amount of input length.

The results demonstrate that, for medical tasks where a summarizing label is predicted based on a long, multi-turn conversation, a sentence-level representation is necessary while traditional frame-level feature extraction methods might fail.

\subsection{Ablation}
\label{ssec:ablation}
% We further conduct a series of ablation studies to analyze the different parameter configurations that give rise to our performance gain. Various segment lengths and $k$ along with hop sizes $\omega_{hop}$ are investigated. 
We first design an experiment to demonstrate the importance of response-level representations for depression detection. Then a series of ablation studies are conducted to analyze possible factors which may influence the pretraining model. Lastly, additional generative strategies are compared with the current center spectrogram generation strategy. 

\paragraph{Segment length}
In order to validate the importance of extracting response-level features, we compare the performance of DEPA extracted from different segment lengths (number of
frames) in \Cref{tab:frame}. 
In this experiment, each response is first split into sub-segments of length $L_{seg} \in \{150, 300, 500\}$ ms, whereas for each $L_{seg}$ a single DEPA feature is extracted.
Segments are extracted without overlap, and zero-padding is used for the last segment.
Each segment length can be interpreted as being on a word (150 ms), short sentence (300 ms), and sentence (500 ms) level.

The results indicate an increase in segment length for DEPA extraction commensurates with an increase in binary classification performance.
More significant amounts of frames increase the performance, whereas the performance peaks when DEPA is extracted at the response level (the default).
Note that larger chunks represent a shorter amount of input features the BLSTM needs to process. This subsequently benefits the learning mechanism of the BLSTM. The larger the chunks are, the more information is summarized; hence fewer independent features are extracted.
Interestingly, even though the binary classification performance is greatly enhanced, the MAE score is less influenced. This might indicate that DEPA is an excellent tool for summarizing interviewee sentences yet falls short in extracting meaningful depression-related information.
However, note that an MAE of $\approx 5.5$ is in line with most results in previous works (see \Cref{tab:result}).

\begin{table}[htbp]
    \centering
    \begin{tabular}{c|rrr|rr}
    \multicolumn{1}{c|}{} & \multicolumn{3}{c|}{Classification} & \multicolumn{2}{c}{Regression}\\
    \hline
    Duration & Pre & Rec & F1 & MAE & RMSE \\
     \hline\hline
    % 20 & 0.69 & 0.56 & 0.62 & 5.63 & 6.50 \\
    % 50 & 0.77 & 0.64 & 0.70 & 5.66 & 6.52 \\
    % 100 & 0.71 & 0.70 & 0.71 & 5.67 & 6.52 \\
    150 ms & 0.58 & 0.54 & 0.56 & 5.51 & 6.51 \\
    300 ms & 0.68 & 0.69 & 0.69 & 5.63 & 6.37 \\
    500 ms & 0.78 & 0.76 & 0.77 & 5.57 & 6.43 \\
    response & \textbf{0.93} & \textbf{0.96} & \textbf{0.94} & \textbf{4.75} & \textbf{5.73}\\
    \end{tabular}
    \caption{Comparison on performance of DEPA extracted from different segment lengths. The encoder-decoder has been trained on the AD dataset with hyperparameters $k=5,\omega_{hop}=5 ms$.}
%     \caption{Comparison between detection with and without DEPA pretraining, regarding three utlilzed datasets. $\Sigma$ represents the use of all three datasets for DEPA extraction.}
    \label{tab:frame}
\end{table}

\paragraph{Feature frame-shift}
Experiments comparing differences in frame-shift ($\omega_{hop}$) for DEPA extraction in regards to different values of $k$ can be seen in \Cref{tab:conf}.
Note that for all experiments, the window size $\omega_{win}$ is set to four times the frame-shift.

\begin{table}[htbp]
    \centering
    \begin{tabular}{lc|rrr|rr}
    \multicolumn{2}{c|}{} & \multicolumn{3}{c|}{Classification} & \multicolumn{2}{c}{Regression}\\
    \hline
    $k$  & \makecell[c]{$\omega_{hop}$} & Pre & Rec & F1 & MAE & RMSE \\
     \hline\hline
    \multirow{2}{*}{3} & 5 ms & 0.84 & 0.87 & 0.86 & 4.99 & 6.24 \\
    & 10 ms & 0.62 & 0.62 & 0.62 & 5.49 & 6.47 \\
    \hline
    \multirow{2}{*}{4} & 5 ms & 0.91 & 0.89 & 0.90 & 4.86 & 6.02 \\
     & 10 ms & 0.64 & 0.66 & 0.65 & 5.48 & 6.42 \\
    \hline
    \multirow{2}{*}{5} & 5 ms & \textbf{0.93} & \textbf{0.96} & \textbf{0.94} & \textbf{4.75} & \textbf{5.73}\\
     & 10 ms & 0.72 & 0.68 & 0.70 & 5.43 & 6.59\\
    \end{tabular}
    \caption{Comparison between different configurations of $k$ and the hop size $\omega_{hop}$ during pretraining on the AD dataset.}
    \label{tab:conf}
\end{table}

The results indicate that using a smaller $\omega_{hop}$ enhances performance (compare $\omega_{hop}={5,10}$).
A short hop-size is potentially beneficial to detect slight variations in speech, leading to better performance.

A similar improvement in binary classification performance can be observed when increasing the number of adjacent spectrograms fed into the network $k$, culminating at $k=5$.

\begin{figure}
     \centering
     \begin{subfigure}[b]{0.4\textwidth}
         \centering
         \includegraphics[width=\textwidth]{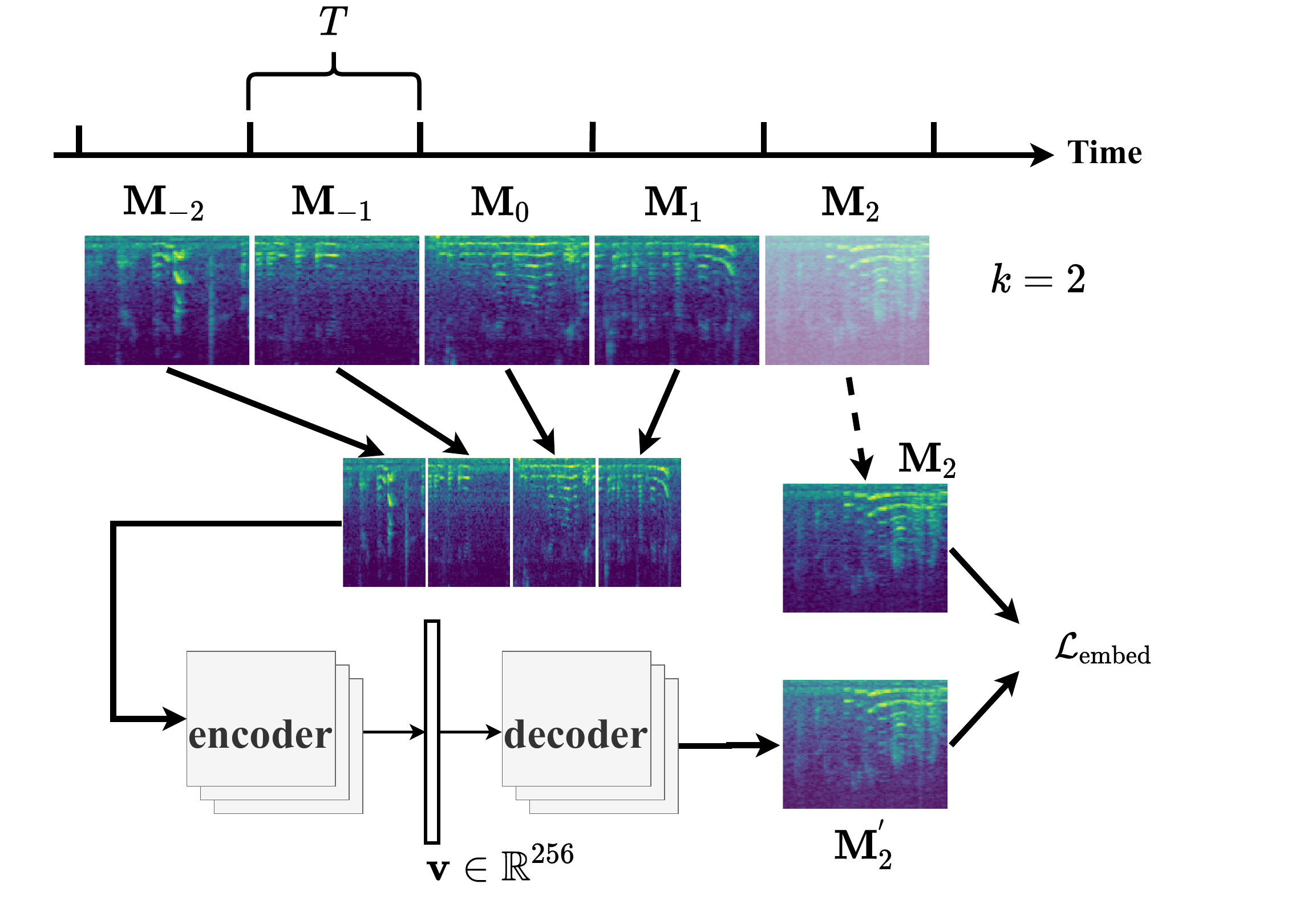}
         \caption{forward strategy}
         \label{fig:forward}
     \end{subfigure}
     \hfill
     \begin{subfigure}[b]{0.4\textwidth}
         \centering
         \includegraphics[width=\textwidth]{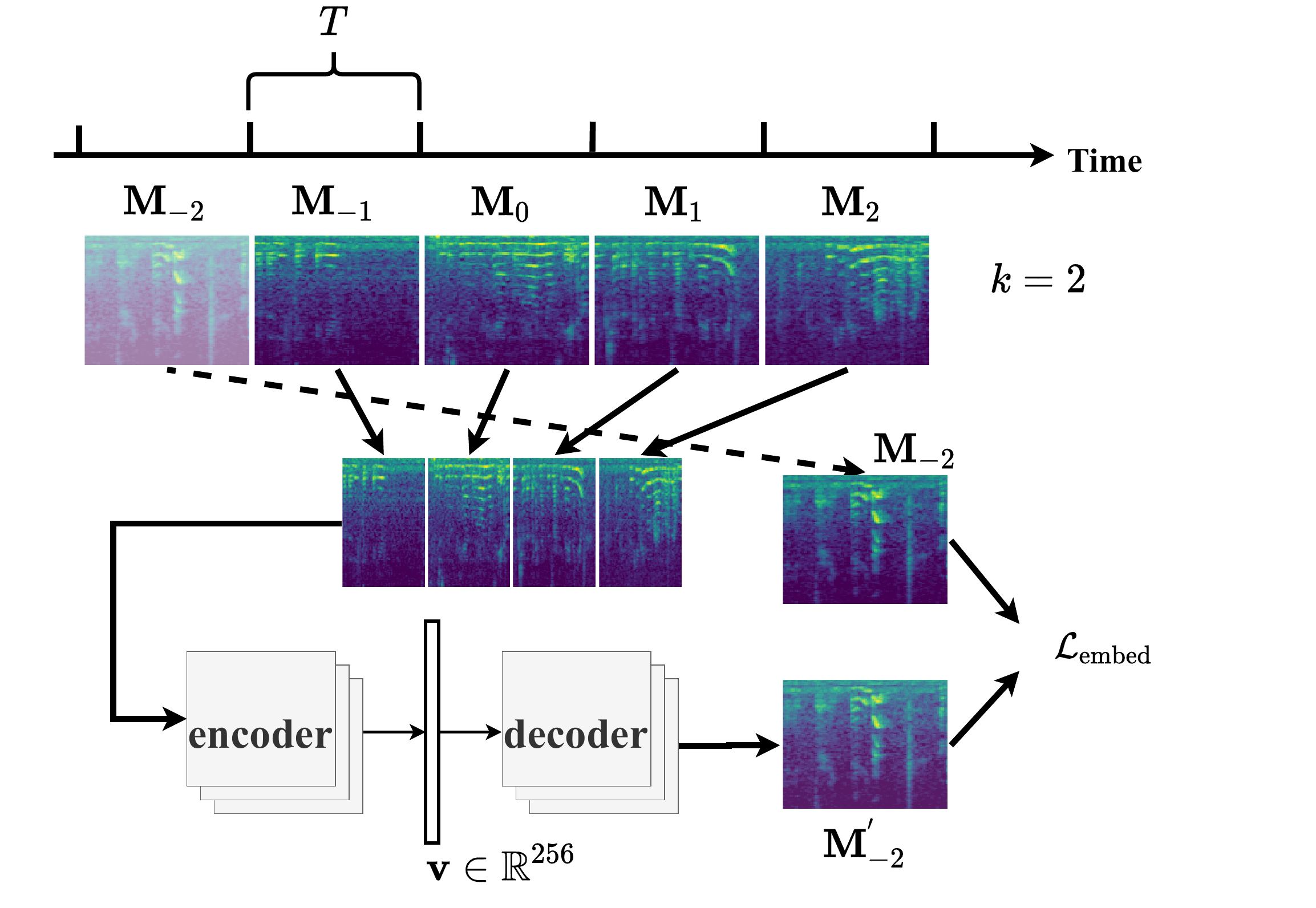}
         \caption{backward strategy}
         \label{fig:backward}
     \end{subfigure}
        \caption{Two different generating strategies}
        \label{fig:patterns}
\end{figure}

\paragraph{Generative Strategy}
In addition to generating the center spectrogram given the context, we conduct experiments to compare the performance of two different generative strategies: forward and backward. In the forward strategy (\Cref{fig:forward}), we predict the last spectrogram $\mathbf{M}_k$ with all previous ones $\mathbf{M}_i,i=[-k,-k+1,\cdots, k-1]$ in the sample, while in the backward strategy (\Cref{fig:backward}), the first spectrogram $\mathbf{M}_{-k}$ is treated as prediction target. Results in \Cref{tab:patterns} indicate that generating the center spectrogram performs the best among three different patterns. It's probably because that the center strategy combines both forward and backward ones and benefits from it. Furthermore, the backward strategy outperforms the forward one, probably because it's harder to figure out the cause (backward) than the result (forward), and a harder pretraining process is beneficial to the downstream task depression detection.

\begin{table}[htbp]
    \centering
    \begin{tabular}{c|rrr|rr}
     & \multicolumn{3}{c|}{Classification} & \multicolumn{2}{c}{Regression}\\
    \hline
    pattern & Pre & Rec & F1 & MAE & RMSE \\
     \hline\hline
    center & \textbf{0.93} & \textbf{0.96} & \textbf{0.94} & \textbf{4.75} & \textbf{5.73} \\
    backward & 0.91 & 0.89 & 0.90 & 5.38 & 6.24 \\
    forward & 0.89 & 0.85 &  0.87 &  5.47 & 6.47 \\
    \end{tabular}
    \caption{Results of different generating patterns}
    \label{tab:patterns}
\end{table}

\section{Conclusion}
This work proposes DEPA, an audio embedding pretraining method for automatic depression detection.
An encoder-decoder model is trained in a self-supervised fashion to predict and reconstruct a center spectrogram given a spectrogram context.
Then, DEPA is extracted from a trained encoder model and fed into a depression detection BLSTM network.
DEPA exhibits an excellent performance compared to traditional LMS, STFT, HCVP, x-vector, and other commonly used features.
DEPA pretrained on \textit{In-domain DAIC} suggests a significantly better result on detection presence detection using STFT features (DAIC F1 0.90, MDD F1 0.94) compared to LMS features (DAIC F1 0.68, MDD F1 0.71) as well as other approaches without DEPA.
Pretraining on large datasets, e.g. DEPA on AD reached F1 0.94 \& MAE 4.75, further shows that additional out-domain data is beneficial to depression detection research.
Results validated on MDD also illustrate performance enhancement with DEPA's superior capability of summarizing sequences.
%Ablation studies verify that the choice of hyperparameters, specifically the frame-shift and the level of DEPA extraction, are crucial for performance.
DEPA can be a generalized method for similar tasks that need to summarize long sequences given a single output label, for e.g. most medical applications of audio classification including dementia, Parkinson's disease.

\section{Acknowledgements}

This work has been supported by National Natural Science Foundation of China (No.61901265), Shanghai Pujiang Program (No.19PJ1406300), State Key Laboratory of Media Convergence Production Technology and Systems Project (No.SKLMCPTS2020003) and Shanghai Municipal Science and Technology Major Project (2021SHZDZX0102). We thank our collaboration with Bio-X Institute, Shanghai Jiao Tong University for the permission to use MDD dataset. Experiments have been carried out on the PI supercomputer at Shanghai Jiao Tong University. Experiments have been carried out on the PI supercomputer at Shanghai Jiao Tong University.

\vfill\eject
\bibliographystyle{ACM-Reference-Format}
\bibliography{reference}

% \appendix









\end{document}